\documentclass[letterpaper,english,aps,prd,reprint]{revtex4-1}
\usepackage[T1]{fontenc}
\usepackage[latin9]{inputenc}
\usepackage{amsmath}
\usepackage{amssymb}

\makeatletter

\pdfpageheight\paperheight
\pdfpagewidth\paperwidth

\usepackage{slashed}
\usepackage{bbm}

\makeatother

\usepackage{babel}
\begin{document}
\title{Little group generators for Dirac neutrino one-particle states}
\author{R. Romero}
\email{rromero@correo.cua.uam.mx}

\address{{\large{}Departamento de Ciencias Naturales, Unidad Cuajimalpa, Universidad
Aut\'onoma Metropolitana,} {\large{}Av. Vasco de Quiroga 4871. Colonia
Santa Fe Cuajimalpa, Alcald\'ia Cuajimalpa de Morelos, Ciudad de M\'exico,
M\'exico 05348.} }
\begin{abstract}
Assuming neutrinos to be of the Dirac type, the little group generators
for the one-particle states, created off the vacuum by the field operator,
are obtained, both in terms of the one-particle states themselves
and in terms of creation/annihilation operators. It is shown that
these generators act also as rotation operators in the Hilbert space
of the states, providing three types of transformations: a helicity
flip, the standard charge conjugation, and a combination of the two,
up to phases. The transformations' properties are provided in detail
and their physical implications discussed. It is also shown that one
of the transformations continues to hold for chiral fields without
mixing them.
\end{abstract}
\maketitle

\section{Introduction}

It is now established that neutrinos oscillate in flavor and are therefore
massive\citep{Fukuda:1998mi,Fukuda:2001nj,Ahn:2002up,Ahmad:2002jz,Eguchi:2002dm,Araki:2004mb,Michael:2006rx},
and one fundamental aspect still unresolved is the determination of
their nature, whether they are Dirac or Majorana particles\citep{Zralek:1997sa,fukugita2003physics,Mohapatra:724618,giunti2007fundamentals,Petcov:2013poa}.
Perturbative calculations do not help in this regard because of the
vanishing small ratio $m/E$ that all differences between the two
types are proportional to\citep{Li:1981um,KAYSER1982137,Kayser:1982br,Mohapatra:724618}.
Experimentally, the observation of neutrinoless double beta decay
processes would confirm their Majorana nature, and there are already
various types of experiments, both planned and underway, set up with
that purpose\citep{Bilenky:2014uka,doi:10.1142/S0217732316300172,DellOro:2016tmg,Vergados:2016hso,Maneschg:2017mzu,DiDomizio:2017rkc}.
However, the non-observation of such processes do not necessarily
imply that neutrinos are of the Dirac type\citep{HIRSCH2018302}.
On the theoretical side, Majorana neutrinos are preferred because
they are central in the various types of the see-saw mechanism\citep{GellMann:1980vs,Yanagida:1979as,Glashow1980,PhysRevLett.44.912,PhysRevD.25.2951,Mohapatra:2006gs,Gouvea:2016shl},
and also in leptogenesis models\citep{Buchmuller:2005eh,doi:10.1080/00107514.2012.701096,Fong:2013wr}.

Associated with the nature of neutrinos is the question of lepton-number
conservation. Let us consider mass eigenstates for both neutrino types.
These are one-particle states of definite energy and momentum created
off the vacuum by the relevant field operator. In terms of creation
operators, $a^{\dagger}\left(\mathbf{p}\right)$ ($b^{\dagger}\left(\mathbf{p}\right)$)
for Dirac neutrinos (anti-neutrinos) and $\hat{a}^{\dagger}\left(\mathbf{p}\right)$
for Majorana neutrinos, the alternatives are, respectively, 

\begin{equation}
a_{-}^{\dagger}\left(\mathbf{p}\right)\neq b_{+}^{\dagger}\left(\mathbf{p}\right),\label{eq:1.1}
\end{equation}

\begin{equation}
\hat{a}_{+}^{\dagger}\left(\mathbf{p}\right)\neq\hat{a}_{-}^{\dagger}\left(\mathbf{p}\right),\label{eq:1.2}
\end{equation}

\noindent where the subscripts $\pm$ respectively denotes positive
and negative helicity. Thus, if lepton number conservation holds,
different helicity neutrinos and anti-neutrinos are different particles
(Dirac case), and Eq. (\ref{eq:1.1}) is the right choice (there is
also $a_{-}^{\dagger}\left(\mathbf{p}\right)\neq b_{+}^{\dagger}\left(\mathbf{p}\right)$,
but these modes have not been observed). On the other hand, if lepton
number is violated neutrinos and anti-neutrinos are just two helicity
states of the same particle (Majorana case), and Eq. (\ref{eq:1.2})
is the right one. Neutrinos and anti-neutrinos are related by the
anti-unitary and discrete $\mathcal{CPT}$ transformation (charge
conjugation, parity and time reversal)\citep{giunti2007fundamentals}.
If $\nu\left(\boldsymbol{p},h\right)$ represents a Dirac neutrino
with momentum $\boldsymbol{p}$ an helicity $h$ then, setting the
overall phase to one, for simplicity, we have

\begin{equation}
\nu\left(\boldsymbol{p},h\right)\xrightarrow{\mathcal{CPT}}\overline{\nu}\left(\boldsymbol{p},-h\right),\label{eq:1.3}
\end{equation}

\noindent where $\overline{\nu}\left(\boldsymbol{p},-h\right)$ is
the anti-neutrino with the same momentum and opposite helicity. For
the Majorana case the $\mathcal{CPT}$ transformation just reverses
the helicity

\begin{equation}
\nu\left(\boldsymbol{p},h\right)\xrightarrow{\mathcal{CPT}}\nu\left(\boldsymbol{p},-h\right).\label{eq:1.4}
\end{equation}

In this work we take Eq. (\ref{eq:1.1}) as the premise and consider
neutrino and anti-neutrino one-particles states, labeled by their
momentum and helicity. That is, we restrict the discussion to the
Hilbert space of the free one-particle states of a given momentum,
created off the vacuum by the relevant field operator, which in the
case of Dirac neutrinos it is four dimensional. Let us, for now, generically
denote them by $\left|\mathbf{p},h\right\rangle $ (they will be properly
defined in the next section). They are degenerate in the four-momentum
eigenvalues and, in particular, if $\mathcal{H}$ is the Hamiltonian
operator we have

\begin{equation}
\mathcal{H}\left|\mathbf{p},h\right\rangle =E_{\boldsymbol{p}}\left|\mathbf{p},h\right\rangle ,\label{eq:1.5}
\end{equation}

\noindent with $E_{\boldsymbol{p}}=\sqrt{\boldsymbol{p}^{2}+m^{2}}$.
Now let us consider a unitary transformation $U$ such that 

\begin{equation}
U\left|\mathbf{p},h\right\rangle =\alpha\left|\mathbf{p},h'\right\rangle ,\label{eq:1.6}
\end{equation}

\noindent where $\alpha$ is a phase, $\left|\alpha\right|^{2}=1$.
That is, $U$ transforms the one-particle states among themselves.
Then it is easy to see that $\left(\mathcal{H}U-U\mathcal{H}\right)\left|\mathbf{p},h\right\rangle =0$.
Thus, 

\begin{equation}
\left[U,\mathcal{H}\right]=0,\label{eq:1.7}
\end{equation}

\noindent and we conclude by Eqs. (\ref{eq:1.5}) and (\ref{eq:1.6})
that $U$ is a unitary transformation leaving the four-momentum invariant,
which is precisely the definition of a little group transformation\citep{10.2307/1968551,tung1985group,cornwell1997group,Costa:2012zz}.
Since there are four one-particle states, two helicity values for
each of the neutrino and the anti-neutrino, there are three different
types of unitary transformations we can consider, these are: a transformation
that flips the helicity without mixing particles and anti-particles,
the standard charge conjugation and a combination of these two. Also,
because the one-particle states are fermionic and massive, the little
group is SU(2), the rotation group for $SL(2,C)$ in the $\left(1/2,0\right)\oplus\left(0,1/2\right)$
representation\citep{tung1985group,Costa:2012zz}. 

It is the purpose of this paper to define the $U$ transformations,
both in terms of the one-particle states themselves and in terms of
creation/annihilation operators, and exhibit their properties, which
are of physical interest. Among other properties, we show that the
transformations are Hermitian besides being unitary, and that they
do indeed satisfy the SU(2) Lie algebra. Physically, the three transformations
correspond to helicity flip, charge conjugation and a combination
of the two, up to phases. This last transformation will be also shown
to hold for chirally projected fields. 

The organization is as follows: We first present the states, operators,
and spinors and establish the conventions in section II, we then proceed
to present and discuss the three unitary transformations in section
III, first in terms of the one-particle states, assuming a finite
volume quantization, and then in terms of creation/annihilation operators,
which is more fundamental. In section IV we show that the Dirac field
operator, both for the unconstrained case and for the left-chiral
one, is consistently transformed under one of the little group transformations.
Finally, we further discuss the physical implications of the results
and provide concluding remarks.

\section{Free field conventions}

Let us assume that a free massive neutrino is of the Dirac type, so
that Eq.(\ref{eq:1.1}) holds. It is thus described by a massive Dirac
field operator, here given in the helicity basis\citep{duncan2012conceptual}

\begin{align}
\begin{split}\Psi(x)= & \int\frac{d^{3}p}{\left(2\pi\right)^{3}}\frac{1}{\sqrt{2E}}\sum_{\lambda=\pm}\left(u_{\lambda}\left(\mathbf{p}\right)a_{\lambda}\left(\mathbf{p}\right)e^{-ip.x}\right.\\
 & +\left.v_{\lambda}\left(\mathbf{p}\right)b_{\lambda}^{\dagger}\left(\mathbf{p}\right)e^{ip.x}\right),
\end{split}
\label{eq:2.1}
\end{align}

\noindent where the operators $a_{\pm}^{\dagger}\left(\mathbf{p}\right)$,
$b_{\pm}^{\dagger}\left(\mathbf{p}\right)$, respectively create particles
and anti-particles of the given helicity, labeled by the subscript,
off the vacuum. The equal-time anti-commutation relations are the
canonical ones\citep{peskin1995introduction}

\begin{equation}
\begin{gathered}\left\{ \Psi_{\alpha}\left(\mathbf{x}\right),\Psi_{\beta}^{\dagger}\left(\mathbf{y}\right)\right\} =\delta^{3}\left(\mathbf{x}-\mathbf{y}\right)\delta_{\alpha\beta},\\
\left\{ \Psi_{\alpha}\left(\mathbf{x}\right),\Psi_{\beta}\left(\mathbf{y}\right)\right\} =\left\{ \Psi_{\alpha}^{\dagger}\left(\mathbf{x}\right),\Psi_{\beta}^{\dagger}\left(\mathbf{y}\right)\right\} =0,\\
\left\{ a_{\lambda}\left(\mathbf{p}\right),a_{\lambda'}^{\dagger}\left(\mathbf{q}\right)\right\} =\left(2\pi\right)^{3}\delta^{3}\left(\mathbf{p}-\mathbf{q}\right)\delta_{\lambda\lambda'}.\\
\left\{ b_{\lambda}\left(\mathbf{p}\right),b_{\lambda'}^{\dagger}\left(\mathbf{q}\right)\right\} =\left(2\pi\right)^{3}\delta^{3}\left(\mathbf{p}-\mathbf{q}\right)\delta_{\lambda\lambda'}
\end{gathered}
\label{eq:2.2}
\end{equation}

\noindent The one-particle sates 

\begin{align}
\begin{split}\left|\mathbf{p},-\right\rangle  & =a_{-}^{\dagger}\left(\mathbf{p}\right)\left|0\right\rangle ,\\
\left|\mathbf{p},+\right\rangle  & =a_{+}^{\dagger}\left(\mathbf{p}\right)\left|0\right\rangle ,\\
\left|\mathbf{\overline{p}},-\right\rangle  & =b_{-}^{\dagger}\left(\mathbf{p}\right)\left|0\right\rangle ,\\
\left|\mathbf{\overline{p}},+\right\rangle  & =b_{+}^{\dagger}\left(\mathbf{p}\right)\left|0\right\rangle ,
\end{split}
\label{eq:2.3}
\end{align}

\noindent correspondingly represent left- and right-handed\footnote{In this work, the terms left-handed (LH) and right-handed (RH) always
refer to helicity $\mp$, respectively.} neutrinos, and left- and right-handed anti-neutrinos, with the anti-particle
states distinguished by an over bar, and $\left|0\right\rangle $
denoting the free vacuum state.

The bispinors in the field expansion in Eq.(\ref{eq:2.1}) are expressed
in terms of the two-component Weyl spinors

\begin{equation}
\begin{array}{cc}
\xi_{+}(\mathbf{p})=\begin{pmatrix}e^{-i\frac{\varphi}{2}}\cos\left(\frac{\theta}{2}\right)\\
e^{i\frac{\varphi}{2}}\sin\left(\frac{\theta}{2}\right)
\end{pmatrix}, & \xi_{-}(\mathbf{p})=\begin{pmatrix}-e^{-i\frac{\varphi}{2}}\sin\left(\frac{\theta}{2}\right)\\
e^{i\frac{\varphi}{2}}\cos\left(\frac{\theta}{2}\right)
\end{pmatrix},\end{array}\label{eq:2.4}
\end{equation}

\noindent which satisfy $\boldsymbol{\sigma}\cdot\mathbf{\hat{p}}\,\xi_{\lambda}(\mathbf{p})=\lambda\xi_{\lambda}(\mathbf{p})$,
$\lambda=\pm$, with three-momentum $\mathbf{\hat{p}}=\mathbf{p}/\left|\mathbf{p}\right|=\left(\sin\theta\cos\varphi,\sin\theta\sin\varphi,\cos\theta\right)$.
Explicitly

\begin{equation}
\begin{split}u_{\lambda}\left(\mathbf{p}\right)= & \begin{pmatrix}\sqrt{E-\lambda\left|\mathbf{p}\right|}\xi_{\lambda}\left(\mathbf{p}\right)\\
\sqrt{E+\lambda\left|\mathbf{p}\right|}\xi_{\lambda}\left(\mathbf{p}\right)
\end{pmatrix},\\
v_{\lambda}\left(\mathbf{p}\right)= & \begin{pmatrix}-\lambda\sqrt{E+\lambda\left|\mathbf{p}\right|}\xi_{-\lambda}\left(\mathbf{p}\right)\\
\lambda\sqrt{E-\lambda\left|\mathbf{p}\right|}\xi_{-\lambda}\left(\mathbf{p}\right)
\end{pmatrix},
\end{split}
\label{eq:2.5}
\end{equation}

\noindent and they are normalized according to the relations

\begin{align}
\begin{split}\overline{u}_{\lambda}\left(\mathbf{p}\right)u_{\lambda'}\left(\mathbf{p}\right)= & 2m\delta_{\lambda,\lambda'}\\
\overline{v}_{\lambda}\left(\mathbf{p}\right)v_{\lambda'}\left(\mathbf{p}\right)= & -2m\delta_{\lambda,\lambda'}\\
\overline{u}_{\lambda}\left(\mathbf{p}\right)v_{\lambda'}\left(\mathbf{p}\right)= & \overline{v}_{\lambda}\left(\mathbf{p}\right)u_{\lambda'}\left(\mathbf{p}\right)=0.
\end{split}
\label{eq:2.6}
\end{align}

\noindent Here, the over bar represents the Dirac adjoint $\overline{u}\equiv u^{\dagger}\gamma^{0}$.
The bispinors satisfy the momentum-space Dirac equations $\left(\slashed{p}-m\right)u_{\lambda}\left(\mathbf{p}\right)=0$
and $\left(\slashed{p}+m\right)v_{\lambda}\left(\mathbf{p}\right)=0$,
with $\slashed{p}\equiv\gamma^{\mu}p_{\mu}$. We use the Weyl representation
of the gamma matrices, as given in Ref. \citep{peskin1995introduction}.

The Hamiltonian, momentum, and lepton-number operators are respectively
given by 

\noindent 
\begin{equation}
\mathcal{H}=\int\frac{d^{3}p}{\left(2\pi\right)^{3}}\sum_{\lambda=\pm}E_{\mathbf{p}}\left(a_{\lambda}^{\dagger}\left(\mathbf{p}\right)a_{\lambda}\left(\mathbf{p}\right)+b_{\lambda}^{\dagger}\left(\mathbf{p}\right)b_{\lambda}\left(\mathbf{p}\right)\right),\label{eq:2.7}
\end{equation}

\begin{equation}
\boldsymbol{P}=\int\frac{d^{3}p}{\left(2\pi\right)^{3}}\sum_{\lambda=\pm}\boldsymbol{p}\left(a_{\lambda}^{\dagger}\left(\mathbf{p}\right)a_{\lambda}\left(\mathbf{p}\right)+b_{\lambda}^{\dagger}\left(\mathbf{p}\right)b_{\lambda}\left(\mathbf{p}\right)\right),\label{eq:2.8}
\end{equation}

\noindent 
\begin{equation}
\mathrm{L}=\int\frac{d^{3}p}{\left(2\pi\right)^{3}}\sum_{\lambda=\pm}\left(a_{\lambda}^{\dagger}\left(\mathbf{p}\right)a_{\lambda}\left(\mathbf{p}\right)-b_{\lambda}^{\dagger}\left(\mathbf{p}\right)b_{\lambda}\left(\mathbf{p}\right)\right).\label{eq:2.9}
\end{equation}

\section{Little group transformations}

\subsection{One-particle states}

To simplify calculations and obtain a rapid overview of the transformations
let us for now resort to a discrete volume quantization\citep{giunti2007fundamentals},
so that the one-particle states in Eq. (\ref{eq:2.3}) can be taken
orthonormal. It is then straightforward to consider operators of the
form $\sum\alpha_{hh'}\left|\mathbf{p},h\right\rangle \left\langle \mathbf{p},h'\right|$,
with $\alpha_{hh'}$ adequately chosen phases so that the operators
are unitary and satisfy the SU(2) algebra. These constitute the $U$
operators in Eqs. (\ref{eq:1.5}) and (\ref{eq:1.6}). Then we have

\begin{equation}
\begin{aligned}\begin{split}U_{1} & =i\left|\mathbf{p},-\right\rangle \left\langle \mathbf{p},+\right|-i\left|\mathbf{p},+\right\rangle \left\langle \mathbf{p},-\right|\\
 & -i\left|\mathbf{\overline{p}},-\right\rangle \left\langle \mathbf{\overline{p}},+\right|+i\left|\mathbf{\overline{p}},+\right\rangle \left\langle \mathbf{\overline{p}},-\right|,
\end{split}
\end{aligned}
\label{eq:3.1}
\end{equation}

\begin{align}
\begin{split}U_{2} & =\left|\mathbf{p},-\right\rangle \left\langle \overline{\mathbf{p}},-\right|+\left|\mathbf{\overline{p}},-\right\rangle \left\langle \mathbf{p},-\right|\\
 & +\left|\mathbf{p},+\right\rangle \left\langle \mathbf{\overline{p}},+\right|+\left|\mathbf{\overline{p}},+\right\rangle \left\langle \mathbf{p},+\right|,
\end{split}
\label{eq:3.2}
\end{align}

\begin{align}
\begin{split}U_{3} & =\left|\mathbf{p},+\right\rangle \left\langle \overline{\mathbf{p}},-\right|-\left|\mathbf{\overline{p}},+\right\rangle \left\langle \mathbf{p},-\right|\\
 & -\left|\mathbf{p},-\right\rangle \left\langle \mathbf{\overline{p}},+\right|+\left|\mathbf{\overline{p}},-\right\rangle \left\langle \mathbf{p},+\right|,
\end{split}
\label{eq:3.3}
\end{align}

\noindent which respectively produce

\begin{equation}
\begin{array}{cc}
\begin{split}U_{1}\left|\mathbf{p},-\right\rangle  & =-i\left|\mathbf{p},+\right\rangle ,\\
U_{1}\left|\mathbf{p},+\right\rangle  & =i\left|\mathbf{p},-\right\rangle ,\\
U_{1}\left|\mathbf{\overline{p}},-\right\rangle  & =i\left|\overline{\mathbf{p}},+\right\rangle ,\\
U_{1}\left|\mathbf{\overline{p}},+\right\rangle  & =-i\left|\mathbf{\overline{p}},-\right\rangle ,
\end{split}
 & \begin{split}\end{split}
\end{array}\label{eq:3.4}
\end{equation}

\begin{equation}
\begin{array}{c}
\begin{split}U_{2}\left|\mathbf{p},-\right\rangle  & =\left|\mathbf{\overline{p}},-\right\rangle ,\\
U_{2}\left|\mathbf{p},+\right\rangle  & =\left|\mathbf{\overline{p}},+\right\rangle ,\\
U_{2}\left|\mathbf{\overline{p}},-\right\rangle  & =\left|\mathbf{p},-\right\rangle ,\\
U_{2}\left|\mathbf{\overline{p}},+\right\rangle  & =\left|\mathbf{p},+\right\rangle ,
\end{split}
\end{array}\label{eq:3.5}
\end{equation}

\begin{equation}
\begin{split}U_{3}\left|\mathbf{p},-\right\rangle  & =-\left|\overline{\mathbf{p}},+\right\rangle ,\\
U_{3}\left|\mathbf{p},+\right\rangle  & =\left|\overline{\mathbf{p}},-\right\rangle ,\\
U_{3}\left|\mathbf{\overline{p}},-\right\rangle  & =\left|\mathbf{p},+\right\rangle ,\\
U_{3}\left|\mathbf{\overline{p}},+\right\rangle  & =-\left|\mathbf{p},-\right\rangle .
\end{split}
\label{eq:3.6}
\end{equation}

To further establish their properties, it is easier to work with a
matrix representation, obtained from the matrix elements 

\begin{equation}
\begin{pmatrix}\left\langle -,\mathbf{p}\right|U_{i}\left|\mathbf{p},-\right\rangle  & \left\langle +,\mathbf{p}\right|U_{i}\left|\mathbf{p},-\right\rangle  & \left\langle -,\overline{\mathbf{p}}\right|U_{i}\left|\mathbf{p},-\right\rangle  & \left\langle +,\overline{\mathbf{p}}\right|U_{i}\left|\mathbf{p},-\right\rangle \\
\left\langle -,\mathbf{p}\right|U_{i}\left|\mathbf{p},+\right\rangle  & \left\langle +,\mathbf{p}\right|U_{i}\left|\mathbf{p},+\right\rangle  & \left\langle -,\overline{\mathbf{p}}\right|U_{i}\left|\mathbf{p},+\right\rangle  & \left\langle +,\overline{\mathbf{p}}\right|U_{i}\left|\mathbf{p},+\right\rangle \\
\left\langle -,\mathbf{p}\right|U_{i}\left|\overline{\mathbf{p}},-\right\rangle  & \left\langle +,\mathbf{p}\right|U_{i}\left|\overline{\mathbf{p}},-\right\rangle  & \left\langle -,\overline{\mathbf{p}}\right|U_{i}\left|\overline{\mathbf{p}},-\right\rangle  & \left\langle +,\overline{\mathbf{p}}\right|U_{i}\left|\overline{\mathbf{p}},-\right\rangle \\
\left\langle -,\mathbf{p}\right|U_{i}\left|\mathbf{\overline{p}},+\right\rangle  & \left\langle +,\mathbf{p}\right|U_{i}\left|\mathbf{\overline{p}},+\right\rangle  & \left\langle -,\overline{\mathbf{p}}\right|U_{i}\left|\mathbf{\overline{p}},+\right\rangle  & \left\langle +,\overline{\mathbf{p}}\right|U_{i}\left|\mathbf{\overline{p}},+\right\rangle 
\end{pmatrix}.\label{eq:3.7}
\end{equation}

\noindent Thus, 

\begin{equation}
U_{1}=\begin{pmatrix}0 & -i & 0 & 0\\
i & 0 & 0 & 0\\
0 & 0 & 0 & i\\
0 & 0 & -i & 0
\end{pmatrix},\label{eq:3.8}
\end{equation}

\begin{equation}
U_{2}=\begin{pmatrix}0 & 0 & 1 & 0\\
0 & 0 & 0 & 1\\
1 & 0 & 0 & 0\\
0 & 1 & 0 & 0
\end{pmatrix},\label{eq:3.9}
\end{equation}

\begin{equation}
U_{3}=\begin{pmatrix}0 & 0 & 0 & -1\\
0 & 0 & 1 & 0\\
0 & 1 & 0 & 0\\
-1 & 0 & 0 & 0
\end{pmatrix}.\label{eq:3.10}
\end{equation}

\noindent Where, with a slight abuse of notation, we label the matrix
representation with the same symbol as the corresponding operator.
With these the following properties are readily verified

\begin{align}
\begin{split}U_{i}^{-1} & =U_{i}^{\dagger},\\
U_{i} & =U_{i}^{\dagger},\\
\det U_{i} & =1,\\
\text{tr}\,U_{i} & =0,\\
\left[\mathcal{H},U_{j}\right] & =0,\\
\left[U_{i},U_{j}\right] & =2i\varepsilon_{ijk}U_{k}.
\end{split}
 & i=1,2,3\label{eq:3.11}
\end{align}

Being both unitary and Hermitian, the transformations are also observables.
The second last property follows from comparing Eqs. (\ref{eq:3.1})
to (\ref{eq:3.3}) with Eq. (\ref{eq:1.6}) and using Eqs. (\ref{eq:1.5})
and (\ref{eq:1.7}). This property verifies that the transformations
leave the four-momentum invariant and are conserved. The last property
establishes that the transformations fulfill the SU(2) algebra, so
they can be identified with the little group generators. In this regard
they are analogous to the Pauli matrices that play the dual role of
being SU(2) generators and $2\pi$ rotation operators for spin $1/2$
particles.

The transformation's physical content is read directly from Eqs. (\ref{eq:3.4})
- (\ref{eq:3.6}): $U_{1}$ flips the helicity without mixing particles
and anti-particles, $U_{2}$ is charge conjugation with the conventional
phases\citep{peskin1995introduction}, and $U_{3}$ is a combination
of the previous two, up to a phase. Thus, $U_{3}$ relates particles
and anti-particles with opposite helicities and, in particular, connects
a LH neutrino one-particle state with a RH anti-neutrino one. 

We also emphasize that, being a little group rotation, the transformation
does not flip the three-momentum, as it would necessarily be the case
for a $\mathcal{CP}$ (charge conjugation and parity) transformation\citep{giunti2007fundamentals,pokorski2000gauge}.
In fact, using the standard transformation properties of the one-particle
states under $\mathcal{CP}T$, we see from Eq. (\ref{eq:3.6}) that,
up to phases, $U_{3}$ produces the same outcome as a $\mathcal{CPT}$
transformation. This, of course, does not mean that these two operators
are equivalent, since the later is a discrete and anti-unitary spacetime
transformation, while the former is a rotation in spin-space. 

We can readily verify the commutation properties between the little
group rotations with $\mathcal{CP}$ and $\mathcal{CPT}$. Setting
phase factors to one for simplicity, e.g., $\mathcal{CP}\left|\mathbf{p},-\right\rangle =\left|-\mathbf{\overline{p}},+\right\rangle $
and $\mathcal{CP}T\left|\mathbf{p},-\right\rangle =\left|\mathbf{\overline{p}},+\right\rangle $,
and remembering we are restricting the discussion to the Hilbert state
of the free one-particle states, we obtain using Eq. (\ref{eq:3.6})

\begin{align}
\begin{split}U_{3}\mathcal{CP}\left|\mathbf{p},-\right\rangle = & -\left|-\mathbf{p},-\right\rangle ,\\
\mathcal{CP}U_{3}\left|\mathbf{p},-\right\rangle = & -\left|-\mathbf{p},-\right\rangle .
\end{split}
\label{eq:3.12}
\end{align}

\noindent Analogously,

\begin{align}
\begin{split}U_{3}\mathcal{CPT}\left|\mathbf{p},-\right\rangle = & -\left|\mathbf{p},-\right\rangle ,\\
\mathcal{CPT}U_{3}\left|\mathbf{p},-\right\rangle = & -\left|\mathbf{p},-\right\rangle .
\end{split}
\label{eq:3.13}
\end{align}

\noindent Hence,

\begin{equation}
\left[U_{3},\mathcal{CP}\right]=\left[U_{3},\mathcal{CPT}\right]=0.\label{eq:3.14}
\end{equation}

\noindent Similar results hold for $U_{1}$ and $U_{2}$, so we conclude

\begin{equation}
\left[U_{i},\mathcal{CP}\right]=\left[U_{i},\mathcal{CPT}\right]=0,\,\,\,i=1,2,3.\label{eq:3.15}
\end{equation}

\subsection{Creation and annihilation operators}

The little group rotations can also be given in terms of creation
and annihilation operators. They are

\begin{align}
\begin{split}U_{1}= & \exp\left(-i\frac{\pi}{2}\right)\exp\left\{ i\frac{\pi}{2}\int\frac{d^{3}p}{\left(2\pi\right)^{3}}\left(a_{+}^{\dagger}\left(\mathbf{p}\right)a_{-}\left(\mathbf{p}\right)\right.\right.\\
 & +\left.\left.a_{-}^{\dagger}\left(\mathbf{p}\right)a_{+}\left(\mathbf{p}\right)+b_{+}^{\dagger}\left(\mathbf{p}\right)b_{-}\left(\mathbf{p}\right)+b_{-}^{\dagger}\left(\mathbf{p}\right)b_{+}\left(\mathbf{p}\right)\right)\right\} \\
\times & \exp\left\{ i\frac{\pi}{2}\int\frac{d^{3}p}{\left(2\pi\right)^{3}}\left(a_{-}^{\dagger}\left(\mathbf{p}\right)a_{-}\left(\mathbf{p}\right)-a_{+}^{\dagger}\left(\mathbf{p}\right)a_{+}\left(\mathbf{p}\right)\right.\right.\\
 & +\left.\left.b_{+}^{\dagger}\left(\mathbf{p}\right)b_{+}\left(\mathbf{p}\right)-b_{-}^{\dagger}\left(\mathbf{p}\right)b_{-}\left(\mathbf{p}\right)\right)\right\} ,
\end{split}
\label{eq:3.16}
\end{align}

\begin{equation}
\begin{split}U_{2}= & \exp\left\{ i\frac{\pi}{2}\int\frac{d^{3}p}{\left(2\pi\right)^{3}}\sum_{\lambda=\pm}\left[b_{\lambda}^{\dagger}\left(\mathbf{p}\right)-a_{\lambda}^{\dagger}\left(\mathbf{p}\right)\right]\right.\\
\times & \left.\left[a_{\lambda}\left(\mathbf{p}\right)-b_{\lambda}\left(\mathbf{p}\right)\right]\right\} ,
\end{split}
\label{eq:3.17}
\end{equation}

\begin{equation}
\begin{split}U_{3}= & \exp\left\{ i\frac{\pi}{2}\int\frac{d^{3}p}{\left(2\pi\right)^{3}}\left(\left[a_{-}^{\dagger}\left(\mathbf{p}\right)+b_{+}^{\dagger}\left(\mathbf{p}\right)\right]\right.\right.\\
 & \times\left[a_{-}\left(\mathbf{p}\right)+b_{+}\left(\mathbf{p}\right)\right]+\left[b_{-}^{\dagger}\left(\mathbf{p}\right)-a_{+}^{\dagger}\left(\mathbf{p}\right)\right]\\
 & \left.\left.\times\left[a_{+}\left(\mathbf{p}\right)-b_{-}\left(\mathbf{p}\right)\right]\right)\right\} ,
\end{split}
\label{eq:3.18}
\end{equation}

\noindent where we have again slightly abused the notation and keep
the same labels for the transformations. Acting on the creation operators
they give

\begin{equation}
\begin{array}{c}
\begin{split}U_{1}a_{-}^{\dagger}\left(\mathbf{p}\right)U_{1}^{\dagger} & =-ia_{+}^{\dagger}\left(\mathbf{p}\right),\\
U_{1}a_{+}^{\dagger}\left(\mathbf{p}\right)U_{1}^{\dagger} & =ia_{-}^{\dagger}\left(\mathbf{p}\right),\\
U_{1}b_{-}^{\dagger}\left(\mathbf{p}\right)U_{1}^{\dagger} & =ib_{+}^{\dagger}\left(\mathbf{p}\right),\\
U_{1}b_{+}^{\dagger}\left(\mathbf{p}\right)U_{1}^{\dagger} & =-ib_{-}^{\dagger}\left(\mathbf{p}\right),
\end{split}
\end{array}\label{eq:3.19}
\end{equation}

\begin{equation}
\begin{array}{c}
\begin{split}U_{2}a_{-}^{\dagger}\left(\mathbf{p}\right)U_{2}^{\dagger} & =b_{-}^{\dagger}\left(\mathbf{p}\right),\\
U_{2}a_{+}^{\dagger}\left(\mathbf{p}\right)U_{2}^{\dagger} & =b_{+}^{\dagger}\left(\mathbf{p}\right),\\
U_{2}b_{-}^{\dagger}\left(\mathbf{p}\right)U_{2}^{\dagger} & =a_{-}^{\dagger}\left(\mathbf{p}\right),\\
U_{2}b_{+}^{\dagger}\left(\mathbf{p}\right)U_{2}^{\dagger} & =a_{+}^{\dagger}\left(\mathbf{p}\right),
\end{split}
\end{array}\label{eq:3.20}
\end{equation}

\begin{equation}
\begin{array}{c}
\begin{split}U_{3}a_{-}^{\dagger}\left(\mathbf{p}\right)U_{3}^{\dagger} & =-b_{+}^{\dagger}\left(\mathbf{p}\right),\\
U_{3}a_{+}^{\dagger}\left(\mathbf{p}\right)U_{3}^{\dagger} & =b_{-}^{\dagger}\left(\mathbf{p}\right),\\
U_{3}b_{-}^{\dagger}\left(\mathbf{p}\right)U_{3}^{\dagger} & =a_{+}^{\dagger}\left(\mathbf{p}\right),\\
U_{3}b_{+}^{\dagger}\left(\mathbf{p}\right)U_{3}^{\dagger} & =-a_{-}^{\dagger}\left(\mathbf{p}\right).
\end{split}
\end{array}\label{eq:3.21}
\end{equation}

\noindent From Eqs. (\ref{eq:3.16}) - (\ref{eq:3.18}) it is clear
that $U_{i}\left|0\right\rangle =\left|0\right\rangle $, $i=1,2,3$,
with $\left|0\right\rangle $ the vacuum state, then the one-particle
state transformations in Eqs. (\ref{eq:3.4}) - (\ref{eq:3.6}) follow
from Eqs. (\ref{eq:3.19}) - (\ref{eq:3.21}). Explicit construction
of $U_{3}$ and its action is provided in the appendix. It is also
straightforward to verify that, as required, the transformations commute
with the Hamiltonian and the momentum operators in Eqs. (\ref{eq:2.7})
and (\ref{eq:2.8})
\begin{equation}
\left[\mathcal{H},U_{i}\right]=[\boldsymbol{P},U_{i}]=0\,\,\,i=1,2,3.\label{eq:2.12}
\end{equation}

\noindent $U_{2}$ and $U_{3}$ anti-commute with the lepton-number
operator in Eq. (\ref{eq:2.9})

\noindent 
\begin{equation}
\left\{ L,U_{i}\right\} =0\,\,\,i=2,3,\label{eq:2.14}
\end{equation}

\noindent while $U_{1}$ commutes with it. It can also be shown that
the anti-commutation relations remain invariant under the transformations
(the case for $U_{3}$ is shown in the appendix).

\begin{equation}
U_{i}\left\{ \Psi_{\alpha}\left(\mathbf{x}\right),\Psi_{\beta}^{\dagger}\left(\mathbf{y}\right)\right\} U_{i}^{\dagger}=\delta^{3}\left(\mathbf{x}-\mathbf{y}\right)\delta_{\alpha\beta},\,\,\,i=1,2,3.\label{eq:2.25}
\end{equation}

\section{Field operator transformation }

In this section we show that the Dirac field operator is consistently
transformed under $U_{3}$, both for the unconstrained field and the
chirally projected one. 

\subsection{Unconstrained field}

The Dirac field in Eq. (\ref{eq:2.1}) is consistently transformed
under $U_{3}$ by appropriately transforming the bispinors. For that
purpose, let us consider the rotation matrix that implements a counterclockwise
rotation by an angle $2\varphi$ around the positive $p_{z}$ axis

\begin{equation}
\mathcal{R}(\boldsymbol{p})=\begin{pmatrix}R_{C} & 0\\
0 & R_{C}
\end{pmatrix}=\exp\left\{ -i\,2\varphi\dfrac{\Sigma^{3}}{2}\right\} ,\label{eq:4.1}
\end{equation}

\noindent with $\Sigma^{3}$ being the third component of $\boldsymbol{\Sigma}=\begin{pmatrix}\boldsymbol{\sigma} & 0\\
0 & \boldsymbol{\sigma}
\end{pmatrix}$, and $R_{C}$ the SU(2) matrix

\begin{equation}
R_{C}=\begin{pmatrix}e^{-i\varphi} & 0\\
0 & e^{i\varphi}
\end{pmatrix}=\exp\left\{ -i\,2\varphi\dfrac{\sigma^{3}}{2}\right\} \label{eq:4.2}
\end{equation}

\noindent yielding

\begin{equation}
R_{C}\xi_{\pm}^{*}(\mathbf{p})=\xi_{\pm}(\mathbf{p})\label{eq:4.3}
\end{equation}

\noindent on the two-component spinors. Analogously, the matrix $\mathcal{R}(\boldsymbol{p})$
in Eq. (\ref{eq:4.1}) produces, for the bispinors

\begin{equation}
\begin{split}\mathcal{R}(\boldsymbol{p})u_{\lambda}^{*}\left(\mathbf{p}\right) & =u_{\lambda}\left(\mathbf{p}\right),\\
\mathcal{R}(\boldsymbol{p})v_{\lambda}^{*}\left(\mathbf{p}\right) & =v_{\lambda}\left(\mathbf{p}\right).
\end{split}
\label{eq:4.4}
\end{equation}

We also make use of the chiral matrix 

\begin{equation}
\gamma^{5}=i\gamma^{0}\gamma^{1}\gamma^{2}\gamma^{3}=\begin{pmatrix}-1 & 0\\
0 & 1
\end{pmatrix},\label{eq:4.5}
\end{equation}

\noindent and the relations

\begin{equation}
\begin{split}-\gamma^{5}u_{\lambda}\left(\mathbf{p}\right) & =\lambda\,v_{-\lambda}\left(\mathbf{p}\right),\\
-\gamma^{5}v_{\lambda}\left(\mathbf{p}\right) & =-\lambda\,u_{-\lambda}\left(\mathbf{p}\right).
\end{split}
\label{eq:4.6}
\end{equation}

\noindent Combining Eqs. (\ref{eq:4.4}) and (\ref{eq:4.6}) we get

\begin{equation}
\begin{split}-\gamma^{5}\mathcal{R}(\boldsymbol{p})u_{\lambda}^{*}\left(\mathbf{p}\right) & =\lambda\,v_{-\lambda}\left(\mathbf{p}\right),\\
-\gamma^{5}\mathcal{R}(\boldsymbol{p})v_{\lambda}^{*}\left(\mathbf{p}\right) & =-\lambda\,u_{-\lambda}\left(\mathbf{p}\right).
\end{split}
\label{eq:4.7}
\end{equation}

\noindent The field transformation is then obtained from Eqs. (\ref{eq:2.1}),
(\ref{eq:3.21}) and (\ref{eq:4.7}) as

\begin{equation}
U_{3}\Psi(x)U_{3}^{\dagger}=-\gamma^{5}\mathcal{R}(\boldsymbol{p})\Psi^{*}(x),\label{eq:4.8}
\end{equation}

\noindent and it constitutes a consistent transformation of the field
operator, since its rhs induces a transformation of the Dirac equation
in momentum-space: from $-\gamma^{5}\mathcal{R}(\boldsymbol{p})\left(\slashed{p}^{*}-m\right)\left(-\gamma^{5}\mathcal{R}(\boldsymbol{p})\right)^{\dagger}=-\left(\slashed{p}+m\right)$
and Eq. (\ref{eq:4.7}) we get

\begin{equation}
\begin{array}{ccc}
\left(\slashed{p}-m\right)u_{\lambda}\left(\mathbf{p}\right)=0 & {\displaystyle \xrightarrow{-\gamma^{5}\mathcal{R}(\boldsymbol{p})\mathcal{K}}} & \left(\slashed{p}+m\right)v_{-\lambda}\left(\mathbf{p}\right)=0\end{array},\label{eq:4.9}
\end{equation}

\noindent where $\mathcal{K}$ represents the operation of conjugating
to the right. We also have that a second application of the $U_{3}$
transformation corresponds to no transformation at all, as can be
seen from Eq. (\ref{eq:3.21}) and the fact that $\left(-\gamma^{5}\mathcal{R}(\boldsymbol{p})\right)\mathcal{K}\left(-\gamma^{5}\mathcal{R}(\boldsymbol{p})\right)\mathcal{K}=\left(-\gamma^{5}\mathcal{R}(\boldsymbol{p})\right)\left(-\gamma^{5}\mathcal{R}(\boldsymbol{p})\right)^{*}=\boldsymbol{1}$. 

The case for $U_{2}$, being the charge conjugation operator, is textbook
matter, and the case for $U_{1}$ proceeds in a similar fashion.

\subsection{Chirally projected field}

Let us apply the chiral projection operators $L=1/2\left(\boldsymbol{1}-\gamma^{5}\right)$
and $R=1/2\left(\boldsymbol{1}+\gamma^{5}\right)$ to the field expansion
in Eq. (\ref{eq:2.1}), to obtain

\begin{align}
\begin{split}\Psi_{L}(x)= & \int\frac{d^{3}p}{\left(2\pi\right)^{3}}\frac{1}{\sqrt{2E}}\left\{ \left(a_{+}\left(\mathbf{p}\right)e^{-ip.x}\right.\right.\\
 & +\left.b_{-}^{\dagger}\left(\mathbf{p}\right)e^{ip.x}\right)\xi_{+}\left(\mathbf{p}\right)\sqrt{E-\left|\mathbf{p}\right|}+\left(a_{-}\left(\mathbf{p}\right)e^{-ip.x}\right.\\
 & -\left.\left.b_{+}^{\dagger}\left(\mathbf{p}\right)e^{ip.x}\right)\xi_{-}\left(\mathbf{p}\right)\sqrt{E+\left|\mathbf{p}\right|}\right\} ,
\end{split}
\label{eq:4.10}
\end{align}

\begin{align}
\begin{split}\Psi_{R}(x)= & \int\frac{d^{3}p}{\left(2\pi\right)^{3}}\frac{1}{\sqrt{2E}}\left\{ \left(a_{+}\left(\mathbf{p}\right)e^{-ip.x}\right.\right.\\
 & -\left.b_{-}^{\dagger}\left(\mathbf{p}\right)e^{ip.x}\right)\xi_{+}\left(\mathbf{p}\right)\sqrt{E+\left|\mathbf{p}\right|}+\left(a_{-}\left(\mathbf{p}\right)e^{-ip.x}\right.\\
 & +\left.\left.b_{+}^{\dagger}\left(\mathbf{p}\right)e^{ip.x}\right)\xi_{-}\left(\mathbf{p}\right)\sqrt{E-\left|\mathbf{p}\right|}\right\} .
\end{split}
\label{eq:4.11}
\end{align}

\noindent We thus see that each chiral field and its Hermitian conjugate
produce both helicity states, and we can continue to use the one-particle
states in Eq.(\ref{eq:2.3}), with one set for each chirality. 

In the high energy limit, with $E\gg m$, states with weight $\sqrt{E-\left|\mathbf{p}\right|}\approx m/\sqrt{2E}$
are suppressed, while the ones with $\sqrt{E+\left|\mathbf{p}\right|}\approx\sqrt{2E}$
are favored, so in general, and if helicity is not measured, a left-chiral
neutrino of energy $E$ produced by the weak interaction will be in
a superposition of both helicities. It can then be described by the
density matrix\citep{Zralek:1997sa}

\begin{align}
\begin{split}\rho_{\nu}(E)= & \left(\dfrac{E+\left|\mathbf{p}\right|}{2E}\right)\left|\mathbf{p},-\right\rangle \left\langle \mathbf{p},-\right|\\
+ & \left(\dfrac{E-\left|\mathbf{p}\right|}{2E}\right)\left|\mathbf{p},+\right\rangle \left\langle \mathbf{p},+\right|.
\end{split}
\label{eq:4.12}
\end{align}

\noindent The anti-neutrino density matrix is accordingly given by

\begin{align}
\begin{split}\rho_{\bar{\nu}}(E)= & \left(\dfrac{E+\left|\mathbf{p}\right|}{2E}\right)\left|\mathbf{\bar{p}},+\right\rangle \left\langle \mathbf{\bar{p}},+\right|\\
+ & \left(\dfrac{E-\left|\mathbf{p}\right|}{2E}\right)\left|\mathbf{\bar{p}},-\right\rangle \left\langle \mathbf{\bar{p}},-\right|,
\end{split}
\label{eq:4.13}
\end{align}

\noindent and from Eq. (\ref{eq:3.6}) we have

\begin{equation}
U_{3}\rho_{\nu}(E)U_{3}^{\dagger}=\rho_{\bar{\nu}}(E),\label{eq:4.14}
\end{equation}

\noindent and in particular we can again conclude that a LH neutrino
and a RH anti-neutrino, created by a left-chiral field, are connected
by the $U_{3}$ transformation. 

For the field operator in Eq.(\ref{eq:4.10}) we get, using Eqs.(\ref{eq:3.21})
and (\ref{eq:4.3})

\begin{equation}
U_{3}\Psi_{L}(x)U_{3}^{\dagger}=R_{C}\Psi_{L}^{*}(x),\label{eq:4.15}
\end{equation}

\noindent so the left-chiral field transforms appropriately under
$U_{3}$. On the other hand, $U_{2}$ mixes the chiral fields, as
it must for charge conjugation

\begin{equation}
U_{2}\Psi_{L}(x)U_{2}^{\dagger}=-i\sigma_{2}\Psi_{R}^{*}(x).\label{eq:4.16}
\end{equation}

\noindent As for $U_{1}$ it is not possible to obtain a consistent
transformation of the fields and at the same time maintain the SU(2)
algebra, so this transformation is lost for chiral fields.

\section{Concluding remarks}

We have obtained the little group generators, which act also as symmetry
operators, for massive Dirac neutrino one-particle states, provided
their properties in detail and discuss their physical interpretations.
The most interesting result comes from $U_{3}$ because it connects
a LH neutrino state with a RH anti-neutrino one, by a rotation in
spin space that violates lepton number conservation. The other two
transformations involve states that have not been observed, namely
a RH-neutrino and a LH anti-neutrino, but which are in principle not
precluded by any fundamental consideration. Regarding $U_{2}$, which
as stated is just the standard charge conjugation operator, what we
have obtained is consistent with the fact that this transformation
is actually an internal transformation\citep{Weinberg1995}, not related
to spacetime at all.

Let us consider $U_{2}$ and $U_{3}$ for free charged fermions. In
this case charge conjugation is also a symmetry of the free theory
since it commutes with the Hamiltonian, but of course this does not
imply that a charged fermion can spontaneously change to its anti-fermion,
since such a process is precluded by the charge conservation selection
rule. The same applies for $U_{3}$. No such selection rule exists
for strictly neutral fermions so, to the extent that the transformations
here presented are superseded by charge conservation, these apply
only to strictly neutral, elementary fermions, of which the neutrino
is the only particle known to exist so far, in the free theory. The
elementary part is guaranteed by the fundamental aspect of the field
and the use of one-particle states created off the vacuum by the field
operator. 

As for lepton number, even though total lepton number has never been
observed to be violated, it is a classical global symmetry, and there
is a priori no reason, either from unitarity, renormalizability, or
otherwise, that prevents it to be broken by quantum effects. Flavor
lepton number, on the other hand, is already known to be violated
by the flavor basis in neutrino oscillations. 

\appendix

\section{Realization of the $U_{3}$ transformation}

In this appendix we explicitly derive Eq. (\ref{eq:3.21}). Let us
consider the unitary transformations

\begin{align}
\begin{split}\hat{\omega}_{1}= & \exp\left(i\alpha A\right)\\
\hat{\omega}_{2}= & \exp\left(i\beta B\right)
\end{split}
\label{eq:1-1}
\end{align}

\noindent with $\alpha$ and $\beta$ real numbers, and $A$ and $B$
the Hermitian operators 

\begin{align}
\begin{split}A= & \int\frac{d^{3}p}{\left(2\pi\right)^{3}}\left\{ a_{-}^{\dagger}\left(\mathbf{p}\right)a_{-}\left(\mathbf{p}\right)-a_{+}^{\dagger}\left(\mathbf{p}\right)a_{+}\left(\mathbf{p}\right)\right.\\
 & \left.+b_{+}^{\dagger}\left(\mathbf{p}\right)b_{+}\left(\mathbf{p}\right)-b_{-}^{\dagger}\left(\mathbf{p}\right)b_{-}\left(\mathbf{p}\right)\right\} ,
\end{split}
\label{eq:2-1}
\end{align}

\begin{align}
\begin{split}B= & \int\frac{d^{3}p}{\left(2\pi\right)^{3}}\left\{ a_{-}^{\dagger}\left(\mathbf{p}\right)b_{+}\left(\mathbf{p}\right)+b_{+}^{\dagger}\left(\mathbf{p}\right)a_{-}\left(\mathbf{p}\right)\right.\\
 & \left.+b_{-}^{\dagger}\left(\mathbf{p}\right)a_{+}\left(\mathbf{p}\right)+a_{+}^{\dagger}\left(\mathbf{p}\right)b_{-}\left(\mathbf{p}\right)\right\} .
\end{split}
\label{eq:3-1}
\end{align}

\noindent Then, using the operator identity $\left[\hat{A}\hat{B},\hat{C}\right]=\hat{A}\left\{ \hat{B},\hat{C}\right\} -\left\{ \hat{A},\hat{C}\right\} \hat{B}$,
we have $\left[A,a_{-}\left(\mathbf{p}\right)\right]=-a_{-}\left(\mathbf{p}\right)$,
$\left[A,\left[A,a_{-}\left(\mathbf{p}\right)\right]\right]=a_{-}\left(\mathbf{p}\right)$,
and so on recursively. Thus, by the Baker\textendash Campbell\textendash Hausdorff
relation we get $\hat{\omega}_{1}a_{-}\left(\mathbf{p}\right)\hat{\omega}_{1}^{\dagger}=a_{-}\left(\mathbf{p}\right)e^{-i\alpha}$.
Performing analogous calculations for the rest of the operators, and
choosing $\alpha=\frac{\pi}{2}$, yields

\begin{align}
\begin{split}\hat{\omega}_{1}a_{-}\left(\mathbf{p}\right)\hat{\omega}_{1}^{\dagger}= & -ia_{-}\left(\mathbf{p}\right),\\
\hat{\omega}_{1}a_{+}\left(\mathbf{p}\right)\hat{\omega}_{1}^{\dagger}= & ia_{+}\left(\mathbf{p}\right),\\
\hat{\omega}_{1}b_{-}\left(\mathbf{p}\right)\hat{\omega}_{1}^{\dagger}= & ib_{-}\left(\mathbf{p}\right),\\
\hat{\omega}_{1}b_{+}\left(\mathbf{p}\right)\hat{\omega}_{1}^{\dagger}= & -ib_{+}\left(\mathbf{p}\right).
\end{split}
\label{eq:4-1}
\end{align}

As for the $B$ operator we get $\left[B,a_{-}\left(\mathbf{p}\right)\right]=-b_{+}\left(\mathbf{p}\right)$,
$\left[B,\left[B,a_{-}\left(\mathbf{p}\right)\right]\right]=a_{-}\left(\mathbf{p}\right)$,
and so on recursively, yielding $\hat{\omega}_{2}a_{-}\left(\mathbf{p}\right)\hat{\omega}_{2}^{\dagger}=a_{-}\left(\mathbf{p}\right)\cos\beta-ib_{+}\left(\mathbf{p}\right)\sin\beta$.
Performing analogous calculations for the rest of the operators, and
choosing $\beta=\frac{\pi}{2}$, result in 

\begin{equation}
\begin{split}\hat{\omega}_{2}a_{-}\left(\mathbf{p}\right)\hat{\omega}_{2}^{\dagger}= & -ib_{+}\left(\mathbf{p}\right),\\
\hat{\omega}_{2}a_{+}\left(\mathbf{p}\right)\hat{\omega}_{2}^{\dagger}= & -ib_{-}\left(\mathbf{p}\right),\\
\hat{\omega}_{2}b_{-}\left(\mathbf{p}\right)\hat{\omega}_{2}^{\dagger}= & -ia_{+}\left(\mathbf{p}\right),\\
\hat{\omega}_{2}b_{+}\left(\mathbf{p}\right)\hat{\omega}_{2}^{\dagger}= & -ia_{-}\left(\mathbf{p}\right).
\end{split}
\label{eq:5-1}
\end{equation}

\noindent Now, with the help of the identity $\left[\hat{A}\hat{B},\hat{C}\hat{D}\right]=\hat{A}\left\{ \hat{B},\hat{C}\right\} \hat{D}-\left\{ \hat{A},\hat{C}\right\} \hat{B}\hat{D}+\hat{C}\hat{A}\left\{ \hat{B},\hat{D}\right\} -\hat{C}\left\{ \hat{A},\hat{D}\right\} \hat{B}$
it is straightforward to check that the operators $A$ and $B$ commute
and so, with $\alpha=\beta=\frac{\pi}{2}$, we have $\hat{\omega}_{1}\hat{\omega}_{2}=\exp\left\{ i\frac{\pi}{2}\left(A+B\right)\right\} =U_{3}$,
with $U_{3}$ given in Eq. (\ref{eq:3.18}), and where the last equality
is directly verified after factorizing. Thus, combining Eqs. (\ref{eq:4-1})
and (\ref{eq:5-1}) and taking the Hermitian conjugate, Eq. (\ref{eq:3.21})
follows directly.

To check the invariance of the equal-time anti-commutation relations
let us define the unitary matrix 

\begin{equation}
\Gamma=i\gamma^{2}\exp\left\{ i\frac{\pi}{2}\left[\hat{\boldsymbol{n}}\cdot\boldsymbol{\Sigma}^{*}\right]\right\} ,\label{eq:6-1}
\end{equation}

\noindent with $\hat{\boldsymbol{n}}=\left(-\sin\varphi,\cos\varphi,0\right)$.
Using Eqs. (\ref{eq:4.1}) and (\ref{eq:4.2}) it is straightforward to check that $\Gamma\gamma^{0}=-\gamma^{5}\mathcal{R}(\boldsymbol{p})$.
Then the rhs of Eq. (\ref{eq:4.8}) is rewritten as

\begin{equation}
U_{3}\Psi(x)U_{3}^{\dagger}=\Gamma\bar{\Psi}^{T}(x).\label{eq:7-1}
\end{equation}

\noindent Thus,

\begin{align}
\begin{split}U_{3}\left\{ \Psi_{\alpha}\left(\mathbf{x}\right),\Psi_{\beta}^{\dagger}\left(\mathbf{y}\right)\right\} U_{3}^{\dagger}= & \left\{ \left(\Gamma\bar{\Psi}^{T}(x)\right)_{\alpha},\left(\Psi^{T}(x)\gamma^{0}\Gamma^{\dagger}\right)_{\beta}\right\} \\
= & \Gamma_{\alpha\mu}\gamma_{\nu\mu}^{0}\left\{ \Psi_{\nu}\left(\mathbf{x}\right),\Psi_{\sigma}^{\dagger}\left(\mathbf{y}\right)\right\} \gamma_{\sigma\tau}^{0}\Gamma_{\tau\beta}^{\dagger}\\
= & \delta^{3}\left(\mathbf{x}-\mathbf{y}\right)\left(\Gamma\gamma^{0}\gamma^{0}\Gamma^{\dagger}\right)_{\alpha\beta}\\
= & \delta^{3}\left(\mathbf{x}-\mathbf{y}\right)\delta_{\alpha\beta}.
\end{split}
\label{eq:8-1}
\end{align}

\bibliographystyle{apsrev4-2}

\begin{thebibliography}{41}%
\makeatletter
\providecommand \@ifxundefined [1]{%
 \@ifx{#1\undefined}
}%
\providecommand \@ifnum [1]{%
 \ifnum #1\expandafter \@firstoftwo
 \else \expandafter \@secondoftwo
 \fi
}%
\providecommand \@ifx [1]{%
 \ifx #1\expandafter \@firstoftwo
 \else \expandafter \@secondoftwo
 \fi
}%
\providecommand \natexlab [1]{#1}%
\providecommand \enquote  [1]{``#1''}%
\providecommand \bibnamefont  [1]{#1}%
\providecommand \bibfnamefont [1]{#1}%
\providecommand \citenamefont [1]{#1}%
\providecommand \href@noop [0]{\@secondoftwo}%
\providecommand \href [0]{\begingroup \@sanitize@url \@href}%
\providecommand \@href[1]{\@@startlink{#1}\@@href}%
\providecommand \@@href[1]{\endgroup#1\@@endlink}%
\providecommand \@sanitize@url [0]{\catcode `\\12\catcode `\$12\catcode
  `\&12\catcode `\#12\catcode `\^12\catcode `\_12\catcode `\%12\relax}%
\providecommand \@@startlink[1]{}%
\providecommand \@@endlink[0]{}%
\providecommand \url  [0]{\begingroup\@sanitize@url \@url }%
\providecommand \@url [1]{\endgroup\@href {#1}{\urlprefix }}%
\providecommand \urlprefix  [0]{URL }%
\providecommand \Eprint [0]{\href }%
\providecommand \doibase [0]{https://doi.org/}%
\providecommand \selectlanguage [0]{\@gobble}%
\providecommand \bibinfo  [0]{\@secondoftwo}%
\providecommand \bibfield  [0]{\@secondoftwo}%
\providecommand \translation [1]{[#1]}%
\providecommand \BibitemOpen [0]{}%
\providecommand \bibitemStop [0]{}%
\providecommand \bibitemNoStop [0]{.\EOS\space}%
\providecommand \EOS [0]{\spacefactor3000\relax}%
\providecommand \BibitemShut  [1]{\csname bibitem#1\endcsname}%
\let\auto@bib@innerbib\@empty
\bibitem [{\citenamefont {Fukuda}\ \emph {et~al.}(1998)\citenamefont {Fukuda}
  \emph {et~al.}}]{Fukuda:1998mi}%
  \BibitemOpen
  \bibfield  {author} {\bibinfo {author} {\bibfnamefont {Y.}~\bibnamefont
  {Fukuda}} \emph {et~al.} (\bibinfo {collaboration} {Super-Kamiokande}),\
  }\href {https://doi.org/10.1103/PhysRevLett.81.1562} {\bibfield  {journal}
  {\bibinfo  {journal} {Phys. Rev. Lett.}\ }\textbf {\bibinfo {volume} {81}},\
  \bibinfo {pages} {1562} (\bibinfo {year} {1998})},\ \Eprint
  {https://arxiv.org/abs/hep-ex/9807003} {arXiv:hep-ex/9807003 [hep-ex]}
  \BibitemShut {NoStop}%
\bibitem [{\citenamefont {Fukuda}\ \emph {et~al.}(2001)\citenamefont {Fukuda}
  \emph {et~al.}}]{Fukuda:2001nj}%
  \BibitemOpen
  \bibfield  {author} {\bibinfo {author} {\bibfnamefont {S.}~\bibnamefont
  {Fukuda}} \emph {et~al.} (\bibinfo {collaboration} {Super-Kamiokande}),\
  }\href {https://doi.org/10.1103/PhysRevLett.86.5651} {\bibfield  {journal}
  {\bibinfo  {journal} {Phys. Rev. Lett.}\ }\textbf {\bibinfo {volume} {86}},\
  \bibinfo {pages} {5651} (\bibinfo {year} {2001})},\ \Eprint
  {https://arxiv.org/abs/hep-ex/0103032} {arXiv:hep-ex/0103032 [hep-ex]}
  \BibitemShut {NoStop}%
\bibitem [{\citenamefont {Ahn}\ \emph {et~al.}(2003)\citenamefont {Ahn} \emph
  {et~al.}}]{Ahn:2002up}%
  \BibitemOpen
  \bibfield  {author} {\bibinfo {author} {\bibfnamefont {M.~H.}\ \bibnamefont
  {Ahn}} \emph {et~al.} (\bibinfo {collaboration} {K2K}),\ }\href
  {https://doi.org/10.1103/PhysRevLett.90.041801} {\bibfield  {journal}
  {\bibinfo  {journal} {Phys. Rev. Lett.}\ }\textbf {\bibinfo {volume} {90}},\
  \bibinfo {pages} {041801} (\bibinfo {year} {2003})},\ \Eprint
  {https://arxiv.org/abs/hep-ex/0212007} {arXiv:hep-ex/0212007 [hep-ex]}
  \BibitemShut {NoStop}%
\bibitem [{\citenamefont {Ahmad}\ \emph {et~al.}(2002)\citenamefont {Ahmad}
  \emph {et~al.}}]{Ahmad:2002jz}%
  \BibitemOpen
  \bibfield  {author} {\bibinfo {author} {\bibfnamefont {Q.~R.}\ \bibnamefont
  {Ahmad}} \emph {et~al.} (\bibinfo {collaboration} {SNO}),\ }\href
  {https://doi.org/10.1103/PhysRevLett.89.011301} {\bibfield  {journal}
  {\bibinfo  {journal} {Phys. Rev. Lett.}\ }\textbf {\bibinfo {volume} {89}},\
  \bibinfo {pages} {011301} (\bibinfo {year} {2002})},\ \Eprint
  {https://arxiv.org/abs/nucl-ex/0204008} {arXiv:nucl-ex/0204008 [nucl-ex]}
  \BibitemShut {NoStop}%
\bibitem [{\citenamefont {Eguchi}\ \emph {et~al.}(2003)\citenamefont {Eguchi}
  \emph {et~al.}}]{Eguchi:2002dm}%
  \BibitemOpen
  \bibfield  {author} {\bibinfo {author} {\bibfnamefont {K.}~\bibnamefont
  {Eguchi}} \emph {et~al.} (\bibinfo {collaboration} {KamLAND}),\ }\href
  {https://doi.org/10.1103/PhysRevLett.90.021802} {\bibfield  {journal}
  {\bibinfo  {journal} {Phys. Rev. Lett.}\ }\textbf {\bibinfo {volume} {90}},\
  \bibinfo {pages} {021802} (\bibinfo {year} {2003})},\ \Eprint
  {https://arxiv.org/abs/hep-ex/0212021} {arXiv:hep-ex/0212021 [hep-ex]}
  \BibitemShut {NoStop}%
\bibitem [{\citenamefont {Araki}\ \emph {et~al.}(2005)\citenamefont {Araki}
  \emph {et~al.}}]{Araki:2004mb}%
  \BibitemOpen
  \bibfield  {author} {\bibinfo {author} {\bibfnamefont {T.}~\bibnamefont
  {Araki}} \emph {et~al.} (\bibinfo {collaboration} {KamLAND}),\ }\href
  {https://doi.org/10.1103/PhysRevLett.94.081801} {\bibfield  {journal}
  {\bibinfo  {journal} {Phys. Rev. Lett.}\ }\textbf {\bibinfo {volume} {94}},\
  \bibinfo {pages} {081801} (\bibinfo {year} {2005})},\ \Eprint
  {https://arxiv.org/abs/hep-ex/0406035} {arXiv:hep-ex/0406035 [hep-ex]}
  \BibitemShut {NoStop}%
\bibitem [{\citenamefont {Michael}\ \emph {et~al.}(2006)\citenamefont {Michael}
  \emph {et~al.}}]{Michael:2006rx}%
  \BibitemOpen
  \bibfield  {author} {\bibinfo {author} {\bibfnamefont {D.~G.}\ \bibnamefont
  {Michael}} \emph {et~al.} (\bibinfo {collaboration} {MINOS}),\ }\href
  {https://doi.org/10.1103/PhysRevLett.97.191801} {\bibfield  {journal}
  {\bibinfo  {journal} {Phys. Rev. Lett.}\ }\textbf {\bibinfo {volume} {97}},\
  \bibinfo {pages} {191801} (\bibinfo {year} {2006})},\ \Eprint
  {https://arxiv.org/abs/hep-ex/0607088} {arXiv:hep-ex/0607088 [hep-ex]}
  \BibitemShut {NoStop}%
\bibitem [{\citenamefont {Zralek}(1997)}]{Zralek:1997sa}%
  \BibitemOpen
  \bibfield  {author} {\bibinfo {author} {\bibfnamefont {M.}~\bibnamefont
  {Zralek}},\ }\href@noop {} {\bibfield  {journal} {\bibinfo  {journal} {Acta
  Phys. Polon.}\ }\textbf {\bibinfo {volume} {B28}},\ \bibinfo {pages} {2225}
  (\bibinfo {year} {1997})},\ \Eprint {https://arxiv.org/abs/hep-ph/9711506}
  {arXiv:hep-ph/9711506 [hep-ph]} \BibitemShut {NoStop}%
\bibitem [{\citenamefont {Fukugita}\ and\ \citenamefont
  {Yanagida}(2003)}]{fukugita2003physics}%
  \BibitemOpen
  \bibfield  {author} {\bibinfo {author} {\bibfnamefont {M.}~\bibnamefont
  {Fukugita}}\ and\ \bibinfo {author} {\bibfnamefont {T.}~\bibnamefont
  {Yanagida}},\ }\href@noop {} {\emph {\bibinfo {title} {Physics of Neutrinos:
  and Application to Astrophysics}}},\ Theoretical and Mathematical Physics\
  (\bibinfo  {publisher} {Springer Berlin Heidelberg},\ \bibinfo {year}
  {2003})\BibitemShut {NoStop}%
\bibitem [{\citenamefont {Mohapatra}\ and\ \citenamefont
  {Pal}(2004)}]{Mohapatra:724618}%
  \BibitemOpen
  \bibfield  {author} {\bibinfo {author} {\bibfnamefont {R.~N.}\ \bibnamefont
  {Mohapatra}}\ and\ \bibinfo {author} {\bibfnamefont {P.~B.}\ \bibnamefont
  {Pal}},\ }\href@noop {} {\emph {\bibinfo {title} {{Massive Neutrinos in
  Physics and Astrophysics; 3rd ed.}}}},\ World Scientific Lecture Notes in
  Physics\ (\bibinfo  {publisher} {World Scientific},\ \bibinfo {address}
  {Singapore},\ \bibinfo {year} {2004})\BibitemShut {NoStop}%
\bibitem [{\citenamefont {Giunti}\ and\ \citenamefont
  {Kim}(2007)}]{giunti2007fundamentals}%
  \BibitemOpen
  \bibfield  {author} {\bibinfo {author} {\bibfnamefont {C.}~\bibnamefont
  {Giunti}}\ and\ \bibinfo {author} {\bibfnamefont {C.}~\bibnamefont {Kim}},\
  }\href@noop {} {\emph {\bibinfo {title} {Fundamentals of Neutrino Physics and
  Astrophysics}}}\ (\bibinfo  {publisher} {OUP, Oxford},\ \bibinfo {year}
  {2007})\BibitemShut {NoStop}%
\bibitem [{\citenamefont {Petcov}(2013)}]{Petcov:2013poa}%
  \BibitemOpen
  \bibfield  {author} {\bibinfo {author} {\bibfnamefont {S.~T.}\ \bibnamefont
  {Petcov}},\ }\href {https://doi.org/10.1155/2013/852987} {\bibfield
  {journal} {\bibinfo  {journal} {Adv. High Energy Phys.}\ }\textbf {\bibinfo
  {volume} {2013}},\ \bibinfo {pages} {852987} (\bibinfo {year} {2013})},\
  \Eprint {https://arxiv.org/abs/1303.5819} {arXiv:1303.5819 [hep-ph]}
  \BibitemShut {NoStop}%
\bibitem [{\citenamefont {Li}\ and\ \citenamefont {Wilczek}(1982)}]{Li:1981um}%
  \BibitemOpen
  \bibfield  {author} {\bibinfo {author} {\bibfnamefont {L.~F.}\ \bibnamefont
  {Li}}\ and\ \bibinfo {author} {\bibfnamefont {F.}~\bibnamefont {Wilczek}},\
  }\href {https://doi.org/10.1103/PhysRevD.25.143} {\bibfield  {journal}
  {\bibinfo  {journal} {Phys. Rev.}\ }\textbf {\bibinfo {volume} {D25}},\
  \bibinfo {pages} {143} (\bibinfo {year} {1982})}\BibitemShut {NoStop}%
\bibitem [{\citenamefont {Kayser}\ and\ \citenamefont
  {Shrock}(1982)}]{KAYSER1982137}%
  \BibitemOpen
  \bibfield  {author} {\bibinfo {author} {\bibfnamefont {B.}~\bibnamefont
  {Kayser}}\ and\ \bibinfo {author} {\bibfnamefont {R.~E.}\ \bibnamefont
  {Shrock}},\ }\href
  {https://doi.org/https://doi.org/10.1016/0370-2693(82)90314-8} {\bibfield
  {journal} {\bibinfo  {journal} {Physics Letters B}\ }\textbf {\bibinfo
  {volume} {112}},\ \bibinfo {pages} {137 } (\bibinfo {year}
  {1982})}\BibitemShut {NoStop}%
\bibitem [{\citenamefont {Kayser}(1982)}]{Kayser:1982br}%
  \BibitemOpen
  \bibfield  {author} {\bibinfo {author} {\bibfnamefont {B.}~\bibnamefont
  {Kayser}},\ }\href {https://doi.org/10.1103/PhysRevD.26.1662} {\bibfield
  {journal} {\bibinfo  {journal} {Phys. Rev.}\ }\textbf {\bibinfo {volume}
  {D26}},\ \bibinfo {pages} {1662} (\bibinfo {year} {1982})}\BibitemShut
  {NoStop}%
\bibitem [{\citenamefont {Bilenky}\ and\ \citenamefont
  {Giunti}(2015)}]{Bilenky:2014uka}%
  \BibitemOpen
  \bibfield  {author} {\bibinfo {author} {\bibfnamefont {S.~M.}\ \bibnamefont
  {Bilenky}}\ and\ \bibinfo {author} {\bibfnamefont {C.}~\bibnamefont
  {Giunti}},\ }\href {https://doi.org/10.1142/S0217751X1530001X} {\bibfield
  {journal} {\bibinfo  {journal} {Int. J. Mod. Phys.}\ }\textbf {\bibinfo
  {volume} {A30}},\ \bibinfo {pages} {1530001} (\bibinfo {year} {2015})},\
  \Eprint {https://arxiv.org/abs/1411.4791} {arXiv:1411.4791 [hep-ph]}
  \BibitemShut {NoStop}%
\bibitem [{\citenamefont {Ostrovskiy}\ and\ \citenamefont
  {O'Sullivan}(2016)}]{doi:10.1142/S0217732316300172}%
  \BibitemOpen
  \bibfield  {author} {\bibinfo {author} {\bibfnamefont {I.}~\bibnamefont
  {Ostrovskiy}}\ and\ \bibinfo {author} {\bibfnamefont {K.}~\bibnamefont
  {O'Sullivan}},\ }\href {https://doi.org/10.1142/S0217732316300172} {\bibfield
   {journal} {\bibinfo  {journal} {Mod. Phys. Lett. A}\ }\textbf {\bibinfo
  {volume} {31}},\ \bibinfo {pages} {1630017} (\bibinfo {year} {2016})},\
  \bibinfo {note} {[Erratum: Mod. Phys. Lett. A 31, no.23, 1692004(2016)]},\
  \Eprint {https://arxiv.org/abs/1605.00631} {arXiv:1605.00631 [hep-ex]}
  \BibitemShut {NoStop}%
\bibitem [{\citenamefont {Dell'Oro}\ \emph {et~al.}(2016)\citenamefont
  {Dell'Oro}, \citenamefont {Marcocci}, \citenamefont {Viel},\ and\
  \citenamefont {Vissani}}]{DellOro:2016tmg}%
  \BibitemOpen
  \bibfield  {author} {\bibinfo {author} {\bibfnamefont {S.}~\bibnamefont
  {Dell'Oro}}, \bibinfo {author} {\bibfnamefont {S.}~\bibnamefont {Marcocci}},
  \bibinfo {author} {\bibfnamefont {M.}~\bibnamefont {Viel}},\ and\ \bibinfo
  {author} {\bibfnamefont {F.}~\bibnamefont {Vissani}},\ }\href
  {https://doi.org/10.1155/2016/2162659} {\bibfield  {journal} {\bibinfo
  {journal} {Adv. High Energy Phys.}\ }\textbf {\bibinfo {volume} {2016}},\
  \bibinfo {pages} {2162659} (\bibinfo {year} {2016})},\ \Eprint
  {https://arxiv.org/abs/1601.07512} {arXiv:1601.07512 [hep-ph]} \BibitemShut
  {NoStop}%
\bibitem [{\citenamefont {Vergados}\ \emph {et~al.}(2016)\citenamefont
  {Vergados}, \citenamefont {Ejiri},\ and\ \citenamefont
  {Simkovic}}]{Vergados:2016hso}%
  \BibitemOpen
  \bibfield  {author} {\bibinfo {author} {\bibfnamefont {J.~D.}\ \bibnamefont
  {Vergados}}, \bibinfo {author} {\bibfnamefont {H.}~\bibnamefont {Ejiri}},\
  and\ \bibinfo {author} {\bibfnamefont {F.}~\bibnamefont {Simkovic}},\ }\href
  {https://doi.org/10.1142/S0218301316300071} {\bibfield  {journal} {\bibinfo
  {journal} {Int. J. Mod. Phys.}\ }\textbf {\bibinfo {volume} {E25}},\ \bibinfo
  {pages} {1630007} (\bibinfo {year} {2016})},\ \Eprint
  {https://arxiv.org/abs/1612.02924} {arXiv:1612.02924 [hep-ph]} \BibitemShut
  {NoStop}%
\bibitem [{\citenamefont {Maneschg}(2017)}]{Maneschg:2017mzu}%
  \BibitemOpen
  \bibfield  {author} {\bibinfo {author} {\bibfnamefont {W.}~\bibnamefont
  {Maneschg}},\ }in\ \href@noop {} {\emph {\bibinfo {booktitle} {{Prospects in
  Neutrino Physics (NuPhys2016) London, United Kingdom, December 12-14,
  2016}}}}\ (\bibinfo {year} {2017})\ \Eprint
  {https://arxiv.org/abs/1704.08537} {arXiv:1704.08537 [physics.ins-det]}
  \BibitemShut {NoStop}%
\bibitem [{\citenamefont {Di}(2017)}]{DiDomizio:2017rkc}%
  \BibitemOpen
  \bibfield  {author} {\bibinfo {author} {\bibfnamefont {S.}~\bibnamefont
  {Di}},\ }in\ \href@noop {} {\emph {\bibinfo {booktitle} {{Prospects in
  Neutrino Physics (NuPhys2016) London, United Kingdom, December 12-14,
  2016}}}}\ (\bibinfo {year} {2017})\ \Eprint
  {https://arxiv.org/abs/1705.03935} {arXiv:1705.03935 [nucl-ex]} \BibitemShut
  {NoStop}%
\bibitem [{\citenamefont {Hirsch}\ \emph {et~al.}(2018)\citenamefont {Hirsch},
  \citenamefont {Srivastava},\ and\ \citenamefont {Valle}}]{HIRSCH2018302}%
  \BibitemOpen
  \bibfield  {author} {\bibinfo {author} {\bibfnamefont {M.}~\bibnamefont
  {Hirsch}}, \bibinfo {author} {\bibfnamefont {R.}~\bibnamefont {Srivastava}},\
  and\ \bibinfo {author} {\bibfnamefont {J.~W.}\ \bibnamefont {Valle}},\ }\href
  {https://doi.org/https://doi.org/10.1016/j.physletb.2018.03.073} {\bibfield
  {journal} {\bibinfo  {journal} {Physics Letters B}\ }\textbf {\bibinfo
  {volume} {781}},\ \bibinfo {pages} {302 } (\bibinfo {year} {2018})},\ \Eprint
  {https://arxiv.org/abs/1711.06181} {arXiv:1711.06181 [hep-ph]} \BibitemShut
  {NoStop}%
\bibitem [{\citenamefont {Gell-Mann}\ \emph {et~al.}(1979)\citenamefont
  {Gell-Mann}, \citenamefont {Ramond},\ and\ \citenamefont
  {Slansky}}]{GellMann:1980vs}%
  \BibitemOpen
  \bibfield  {author} {\bibinfo {author} {\bibfnamefont {M.}~\bibnamefont
  {Gell-Mann}}, \bibinfo {author} {\bibfnamefont {P.}~\bibnamefont {Ramond}},\
  and\ \bibinfo {author} {\bibfnamefont {R.}~\bibnamefont {Slansky}},\ }in\
  \href@noop {} {\emph {\bibinfo {booktitle} {{Supergravity Workshop at Stony
  Brook, New York, September 27-28, 1979}}}},\ Vol.\ \bibinfo {volume}
  {C790927}\ (\bibinfo {year} {1979})\ pp.\ \bibinfo {pages} {315--321},\
  \Eprint {https://arxiv.org/abs/1306.4669} {arXiv:1306.4669 [hep-th]}
  \BibitemShut {NoStop}%
\bibitem [{\citenamefont {Yanagida}(1979)}]{Yanagida:1979as}%
  \BibitemOpen
  \bibfield  {author} {\bibinfo {author} {\bibfnamefont {T.}~\bibnamefont
  {Yanagida}},\ }in\ \href@noop {} {\emph {\bibinfo {booktitle} {{Proceedings
  of the Workshop on the Unified Theories and the Baryon Number in the
  Universe: Tsukuba, Japan, February 13-14, 1979}}}},\ Vol.\ \bibinfo {volume}
  {C7902131}\ (\bibinfo {year} {1979})\ pp.\ \bibinfo {pages}
  {95--99}\BibitemShut {NoStop}%
\bibitem [{\citenamefont {Glashow}(1980)}]{Glashow1980}%
  \BibitemOpen
  \bibfield  {author} {\bibinfo {author} {\bibfnamefont {S.~L.}\ \bibnamefont
  {Glashow}},\ }in\ \href {https://doi.org/10.1007/978-1-4684-7197-7_15} {\emph
  {\bibinfo {booktitle} {Quarks and Leptons: Carg{\`e}se 1979}}},\ \bibinfo
  {editor} {edited by\ \bibinfo {editor} {\bibfnamefont {M.}~\bibnamefont
  {L{\'e}vy}}, \bibinfo {editor} {\bibfnamefont {J.-L.}\ \bibnamefont
  {Basdevant}}, \bibinfo {editor} {\bibfnamefont {D.}~\bibnamefont {Speiser}},
  \bibinfo {editor} {\bibfnamefont {J.}~\bibnamefont {Weyers}}, \bibinfo
  {editor} {\bibfnamefont {R.}~\bibnamefont {Gastmans}},\ and\ \bibinfo
  {editor} {\bibfnamefont {M.}~\bibnamefont {Jacob}}}\ (\bibinfo  {publisher}
  {Springer US},\ \bibinfo {address} {Boston, MA},\ \bibinfo {year} {1980})\
  pp.\ \bibinfo {pages} {687--713}\BibitemShut {NoStop}%
\bibitem [{\citenamefont {Mohapatra}\ and\ \citenamefont
  {Senjanovic}(1980)}]{PhysRevLett.44.912}%
  \BibitemOpen
  \bibfield  {author} {\bibinfo {author} {\bibfnamefont {R.~N.}\ \bibnamefont
  {Mohapatra}}\ and\ \bibinfo {author} {\bibfnamefont {G.}~\bibnamefont
  {Senjanovic}},\ }\href {https://doi.org/10.1103/PhysRevLett.44.912}
  {\bibfield  {journal} {\bibinfo  {journal} {Phys. Rev. Lett.}\ }\textbf
  {\bibinfo {volume} {44}},\ \bibinfo {pages} {912} (\bibinfo {year}
  {1980})}\BibitemShut {NoStop}%
\bibitem [{\citenamefont {Schechter}\ and\ \citenamefont
  {Valle}(1982)}]{PhysRevD.25.2951}%
  \BibitemOpen
  \bibfield  {author} {\bibinfo {author} {\bibfnamefont {J.}~\bibnamefont
  {Schechter}}\ and\ \bibinfo {author} {\bibfnamefont {J.~W.~F.}\ \bibnamefont
  {Valle}},\ }\href {https://doi.org/10.1103/PhysRevD.25.2951} {\bibfield
  {journal} {\bibinfo  {journal} {Phys. Rev. D}\ }\textbf {\bibinfo {volume}
  {25}},\ \bibinfo {pages} {2951} (\bibinfo {year} {1982})}\BibitemShut
  {NoStop}%
\bibitem [{\citenamefont {Mohapatra}\ and\ \citenamefont
  {Smirnov}(2006)}]{Mohapatra:2006gs}%
  \BibitemOpen
  \bibfield  {author} {\bibinfo {author} {\bibfnamefont {R.~N.}\ \bibnamefont
  {Mohapatra}}\ and\ \bibinfo {author} {\bibfnamefont {A.~Y.}\ \bibnamefont
  {Smirnov}},\ }\bibfield  {booktitle} {\emph {\bibinfo {booktitle}
  {{Elementary particle physics. Proceedings, Corfu Summer Institute,
  CORFU2005, Corfu, Greece, September 4-26, 2005}}},\ }\href
  {https://doi.org/10.1146/annurev.nucl.56.080805.140534} {\bibfield  {journal}
  {\bibinfo  {journal} {Ann. Rev. Nucl. Part. Sci.}\ }\textbf {\bibinfo
  {volume} {56}},\ \bibinfo {pages} {569} (\bibinfo {year} {2006})},\ \Eprint
  {https://arxiv.org/abs/hep-ph/0603118} {arXiv:hep-ph/0603118 [hep-ph]}
  \BibitemShut {NoStop}%
\bibitem [{\citenamefont {de~Gouvêa}(2016)}]{Gouvea:2016shl}%
  \BibitemOpen
  \bibfield  {author} {\bibinfo {author} {\bibfnamefont {A.}~\bibnamefont
  {de~Gouvêa}},\ }\href {https://doi.org/10.1146/annurev-nucl-102115-044600}
  {\bibfield  {journal} {\bibinfo  {journal} {Ann. Rev. Nucl. Part. Sci.}\
  }\textbf {\bibinfo {volume} {66}},\ \bibinfo {pages} {197} (\bibinfo {year}
  {2016})}\BibitemShut {NoStop}%
\bibitem [{\citenamefont {Buchmuller}\ \emph {et~al.}(2005)\citenamefont
  {Buchmuller}, \citenamefont {Peccei},\ and\ \citenamefont
  {Yanagida}}]{Buchmuller:2005eh}%
  \BibitemOpen
  \bibfield  {author} {\bibinfo {author} {\bibfnamefont {W.}~\bibnamefont
  {Buchmuller}}, \bibinfo {author} {\bibfnamefont {R.~D.}\ \bibnamefont
  {Peccei}},\ and\ \bibinfo {author} {\bibfnamefont {T.}~\bibnamefont
  {Yanagida}},\ }\href {https://doi.org/10.1146/annurev.nucl.55.090704.151558}
  {\bibfield  {journal} {\bibinfo  {journal} {Ann. Rev. Nucl. Part. Sci.}\
  }\textbf {\bibinfo {volume} {55}},\ \bibinfo {pages} {311} (\bibinfo {year}
  {2005})},\ \Eprint {https://arxiv.org/abs/hep-ph/0502169}
  {arXiv:hep-ph/0502169 [hep-ph]} \BibitemShut {NoStop}%
\bibitem [{\citenamefont {Bari}(2012)}]{doi:10.1080/00107514.2012.701096}%
  \BibitemOpen
  \bibfield  {author} {\bibinfo {author} {\bibfnamefont {P.~D.}\ \bibnamefont
  {Bari}},\ }\href {https://doi.org/10.1080/00107514.2012.701096} {\bibfield
  {journal} {\bibinfo  {journal} {Contemp. Phys.}\ }\textbf {\bibinfo {volume}
  {53}},\ \bibinfo {pages} {315} (\bibinfo {year} {2012})},\ \Eprint
  {https://arxiv.org/abs/1206.3168} {arXiv:1206.3168 [hep-ph]} \BibitemShut
  {NoStop}%
\bibitem [{\citenamefont {Fong}\ \emph {et~al.}(2012)\citenamefont {Fong},
  \citenamefont {Nardi},\ and\ \citenamefont {Riotto}}]{Fong:2013wr}%
  \BibitemOpen
  \bibfield  {author} {\bibinfo {author} {\bibfnamefont {C.~S.}\ \bibnamefont
  {Fong}}, \bibinfo {author} {\bibfnamefont {E.}~\bibnamefont {Nardi}},\ and\
  \bibinfo {author} {\bibfnamefont {A.}~\bibnamefont {Riotto}},\ }\href
  {https://doi.org/10.1155/2012/158303} {\bibfield  {journal} {\bibinfo
  {journal} {Adv. High Energy Phys.}\ }\textbf {\bibinfo {volume} {2012}},\
  \bibinfo {pages} {158303} (\bibinfo {year} {2012})},\ \Eprint
  {https://arxiv.org/abs/1301.3062} {arXiv:1301.3062 [hep-ph]} \BibitemShut
  {NoStop}%
\bibitem [{\citenamefont {Wigner}(1939)}]{10.2307/1968551}%
  \BibitemOpen
  \bibfield  {author} {\bibinfo {author} {\bibfnamefont {E.}~\bibnamefont
  {Wigner}},\ }\href {http://www.jstor.org/stable/1968551} {\bibfield
  {journal} {\bibinfo  {journal} {Annals of Mathematics}\ }\textbf {\bibinfo
  {volume} {40}},\ \bibinfo {pages} {149} (\bibinfo {year} {1939})}\BibitemShut
  {NoStop}%
\bibitem [{\citenamefont {Tung}(1985)}]{tung1985group}%
  \BibitemOpen
  \bibfield  {author} {\bibinfo {author} {\bibfnamefont {W.}~\bibnamefont
  {Tung}},\ }\href@noop {} {\emph {\bibinfo {title} {Group Theory in
  Physics}}}\ (\bibinfo  {publisher} {World Scientific},\ \bibinfo {year}
  {1985})\BibitemShut {NoStop}%
\bibitem [{\citenamefont {Cornwell}(1997)}]{cornwell1997group}%
  \BibitemOpen
  \bibfield  {author} {\bibinfo {author} {\bibfnamefont {J.}~\bibnamefont
  {Cornwell}},\ }\href@noop {} {\emph {\bibinfo {title} {Group Theory in
  Physics: An Introduction}}},\ Techniques of Physics\ (\bibinfo  {publisher}
  {Elsevier Science},\ \bibinfo {year} {1997})\BibitemShut {NoStop}%
\bibitem [{\citenamefont {Costa}\ and\ \citenamefont
  {Fogli}(2012)}]{Costa:2012zz}%
  \BibitemOpen
  \bibfield  {author} {\bibinfo {author} {\bibfnamefont {G.}~\bibnamefont
  {Costa}}\ and\ \bibinfo {author} {\bibfnamefont {G.}~\bibnamefont {Fogli}},\
  }\href {https://doi.org/10.1007/978-3-642-15482-9} {\bibfield  {journal}
  {\bibinfo  {journal} {Lecture Notes in Physics}\ }\textbf {\bibinfo {volume}
  {823}},\ \bibinfo {pages} {1} (\bibinfo {year} {2012})}\BibitemShut {NoStop}%
\bibitem [{\citenamefont {Duncan}(2012)}]{duncan2012conceptual}%
  \BibitemOpen
  \bibfield  {author} {\bibinfo {author} {\bibfnamefont {A.}~\bibnamefont
  {Duncan}},\ }\href@noop {} {\emph {\bibinfo {title} {The Conceptual Framework
  of Quantum Field Theory}}}\ (\bibinfo  {publisher} {Oxford University
  Press},\ \bibinfo {year} {2012})\BibitemShut {NoStop}%
\bibitem [{\citenamefont {Peskin}\ and\ \citenamefont
  {Schroeder}(1995)}]{peskin1995introduction}%
  \BibitemOpen
  \bibfield  {author} {\bibinfo {author} {\bibfnamefont {M.}~\bibnamefont
  {Peskin}}\ and\ \bibinfo {author} {\bibfnamefont {D.}~\bibnamefont
  {Schroeder}},\ }\href@noop {} {\emph {\bibinfo {title} {An introduction to
  quantum field theory}}},\ Advanced book program\ (\bibinfo  {publisher}
  {Addison-Wesley Pub. Co.},\ \bibinfo {year} {1995})\BibitemShut {NoStop}%
\bibitem [{Note1()}]{Note1}%
  \BibitemOpen
  \bibinfo {note} {In this work, the terms left-handed (LH) and right-handed
  (RH) always refer to helicity $\mp $, respectively.}\BibitemShut {Stop}%
\bibitem [{\citenamefont {Pokorski}(2000)}]{pokorski2000gauge}%
  \BibitemOpen
  \bibfield  {author} {\bibinfo {author} {\bibfnamefont {S.}~\bibnamefont
  {Pokorski}},\ }\href@noop {} {\emph {\bibinfo {title} {Gauge Field
  Theories}}},\ Vol.\ \bibinfo {volume} {Cambridge Monographs on Mathematical
  Physics}\ (\bibinfo  {publisher} {Cambridge University Press},\ \bibinfo
  {year} {2000})\BibitemShut {NoStop}%
\bibitem [{\citenamefont {Weinberg}(1995)}]{Weinberg1995}%
  \BibitemOpen
  \bibfield  {author} {\bibinfo {author} {\bibfnamefont {S.}~\bibnamefont
  {Weinberg}},\ }\href@noop {} {\emph {\bibinfo {title} {The Quantum Theory of
  Fields}}},\ \bibinfo {series} {The Quantum Theory of Fields 3 Volume Hardback
  Set}\ No.\ \bibinfo {number} {v. 1}\ (\bibinfo  {publisher} {Cambridge
  University Press},\ \bibinfo {year} {1995})\BibitemShut {NoStop}%
\end{thebibliography}

\end{document}